# Multiple Frequency Steps in Synthetic Antiferromagnet Based Double Spin Josephson Junctions Using CoFeB and Fe$_3$Sn

Md Rakibul Karim Akanda

Department of Engineering Technology, Savannah State University—Savannah, GA 31404, United States of America

**Abstract**
Superconducting quantum interference device (SQUID) which is made of two parallel Josephson junctions has applications in magnetometry. A similar spin-based device is proposed here where spin superfluid in ferromagnet (FM) mimics the superconducting state. Two materials CoFeB and Fe$_3$Sn are used for spin superfluid-based SQUID like device where easy plane anisotropy in CoFeB can be engineered and Fe$_3$Sn has inherent easy plane anisotropy. Frequency varies in spin based proposed devices. Frequency increases and again decreases with the increase in both applied magnetic field and applied spin current. The proposed device can be used as nano oscillator and detector. The frequency in the proposed device shows multiple frequency steps which can be used for neuromorphic applications.'

## 1 Introduction

Superfluidity is the property of a fluid with zero viscosity and is generally assigned to resistance free charge current in a superconductor. Like Cooper pair-based superconductor, superfluidity is also found in spin-based systems where dissipation is negligible and long-distance transport is possible. Spin superfluidity is found in ferrimagnetic material even at room temperature.[1], [2] Spin superfluidity is also found in antiferromagnetic and multiferroic materials.[3], [4], [5]

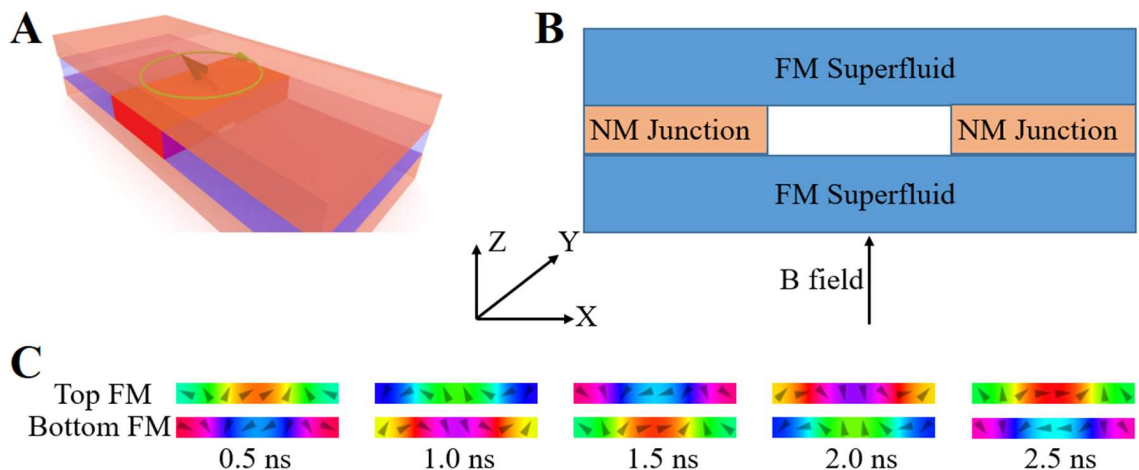

Fig. 1. Two FM is antiferromagnetically exchange coupled by non-magnetic (NM) metal junction at two ends. There is no inter layer exchange coupling in the middle part of top and bottom ferromagnet (FM). Center region is nonmagnetic (NM) insulator. (A) Basic structure showing two parallel Josephson junctions in 3D view where top and bottom ferromagnetic

regions are coupled by violet NM metal junction. Red region is NM insulator where two FM regions are not coupled by spacer layer to make two Josephson junctions at the two ends. (B) Front view of the proposed structure. (C) Change of magnetization in top and bottom FM with time in Fe3Sn with current density of 4 × 108 A/cm2 using MUMAX3.[14]

Apart from using spin Hall effect in the superfluid medium, temperature gradient, laser pulse and domain wall method can also be used.[1], [6], [7] When two Cooper pair-based superconductors are coupled by a weak link e.g. normal metal or insulator, Josephson junction is formed. Conventional superfluidity can be extended as the spin superfluidity in the magnetic system. The magnetic analog of conventional Josephson effect is the spin Josephson effect. This spin Josephson effect can be used as spin nano oscillators where dc electric current or dc magnetic field is converted into magnetization precession.[8], [9], [10] This change in magnetization results in the change in magnetoresistance which can be detected experimentally. These kinds of devices can be implemented as microwave generators or detectors.

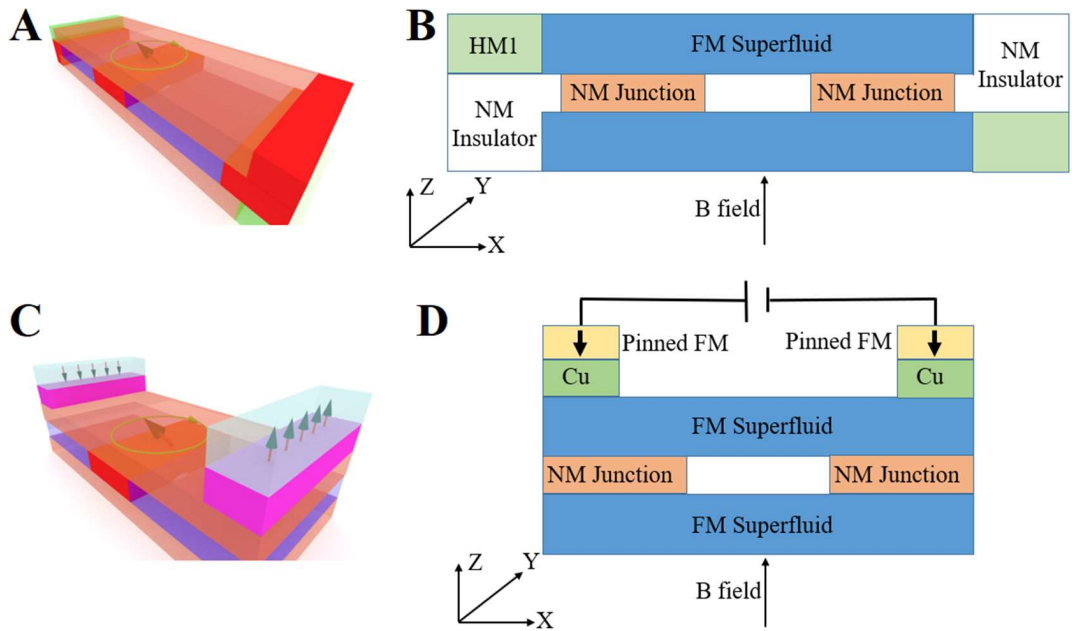

Fig. 2. Current is applied along Y-axis in HM1 and HM2. Current can also be applied in either HM1 or HM2. (A) 3D view at the top where light green region represents heavy metal to inject spin current and (B) side view of the structure. Injected spin current polarized along Z-direction can also be generated using the pinned FM region having copper (Cu) spacer layer and behaving like magnetic tunnel junction (MTJ) or giant magneto-resistor (GMR). (C) 3D view where pink color represents Cu spacer region and (D) Front view.

Dipole effect in the ferromagnets affect the spin superfluid mode and hampers the long-distance transmission.[11], [12] Synthetic antiferromagnet can be used to resolve the problem, where two FM layers are antiferromagnetically coupled through the NM spacer layer. Antiferromagnetic or synthetic antiferromagnetic (SAF) material-based Josephson junction provides terahertz oscillation.[13] Two parallel Josephson junctions form superconducting quantum interference device (SQIUD) which is used for measuring variation of magnetic field as its voltage varies with magnetic field. Like conventional SQIUD, a spin-based device is proposed here which also consists of two parallel junctions. The frequency of this device varies with both magnetic field and current. The variation of frequency shows step-like behavior which can be applicable for neuromorphic applications.[15]-[25]

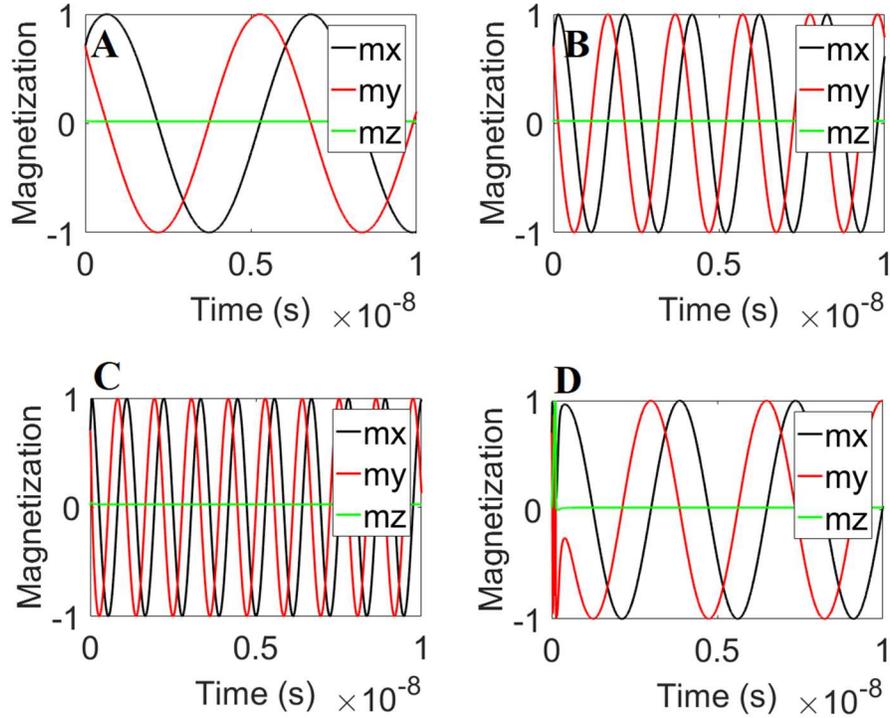

Fig. 3. Variation of spin superfluid oscillation frequency in Fe3Sn with current density (A) $8\times10^7$ A/cm$^2$, (B) $2\times10^8$ A/cm$^2$, (C) $3\times10^8$ A/cm$^2$ and (D) $5\times10^8$ A/cm$^2$. Applied magnetic field is 0.05T along Z-direction. Magnetization along Z-direction is small (but not zero) compared to magnetization along X-direction and Y-direction.

Two ferromagnets are antiferromagnetically exchange coupled by non-magnetic (NM) metal junction at two ends. There is no inter layer exchange coupling in the middle part of top and bottom ferromagnet (FM). White region is nonmagnetic (NM) insulator. Easy plane anisotropy is in X-Y plane. Magnetic fields are applied along Z-axis. Basic structure having two parallel Josephson junctions is shown in Fig. 1. Change of magnetization in top and bottom FM of this structure at different time with applied current is simulated using well known micromagnetic software MUMAX3.[14] Current can be applied along Y-axis in HM1 and HM2 using spin Hall effect (Fig. 2).[26], [27] Current can also be applied in either HM1 or HM2. Charge current, j is along Y-axis. Direction of cross-sectional area (between HM and FM), η is along X-axis. Injected spin current direction due to spin Hall effect, σ (= η × j) is along Z-axis which is necessary for X-Y easy plane anisotropy. Injected spin current polarized along Z-direction can also be generated using the pinned FM region having copper (Cu) spacer layer and behaving like magnetic tunnel junction (MTJ) or giant magneto-resistor (GMR) (Fig. 2).[28], [29] Cu region separates pinned FM and Superfluid region so that superfluid region can move freely. Pinned FM region provides spin current polarized along Z-direction which is necessary for superfluid as easy plane is along XY plane.

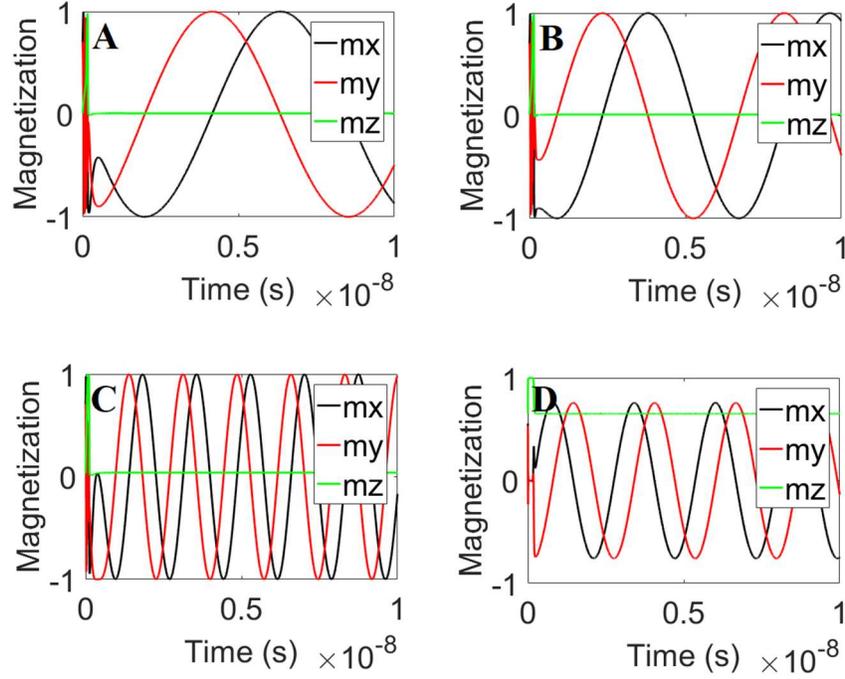

Fig. 4. Variation of spin superfluid oscillation frequency in Fe3Sn with applied magnetic field (A) 0.02T, (B) 0.03T, (C) 0.10T and (D) 2T. Spin current density is $5\times10^8$ A/cm$_2$ which is spin polarized along Z-direction. Magnetization along Z-direction is small (but not zero) compared to magnetization along X-direction or Y-direction and gradually increases with applied magnetic field.

Two types of easy plane ferromagnet CoFeB and Fe$_3$Sn are used where easy plane anisotropy can be engineered in CoFeB whereas Fe$_3$Sn has inherent easy plane anisotropy in X-Y plane. CoFeB is one of the most popular materials for making spintronics based devices.[30]-[32] In the Josephson junctions, top and bottom ferromagnetic layer are antiferromagnetically exchange coupled to provide spin phase difference like SQUID. Fe$_3$Sn has large easy plane anisotropy compared to YIG and is much more suitable for spin superfluid applications. Magnetic field is applied along Z-axis to tilt the magnetization along Z-direction a little bit from original X-Y easy plane. Spin current polarized along Z-direction provides necessary torque for spin superfluid oscillation. Required parameters used in the simulation of SAFM based on CoFeB and Fe$_3$Sn are mentioned in Table I.

|  | CoFeB [33], [34] | Fe$_3$Sn [35], [36] |
| --- | --- | --- |
| Saturation magnetization (Msat) | 0.9 MA/m | 1.18 MA/m |
| Exchange stiffness (Aex) | 14 pJ/m | 10 pJ/m |
| Antiferromagnetic exchange stiffness (Aex) | −1 pJ/m | −1 pJ/m |
| Easy plane anisotropy constant (Ku) | 0.09 MJ/m3 | 1.8 MJ/m3 |
| Landau-Lifshitz damping constant (α) | 0.5 | 1 |
| Easy plane anisotropy | X-Y plane | X-Y plane |

Table I: LLG parameters used in simulations.

## 2. Theory

The spin dynamics of these structures can be described by the Landau-Lifshitz-Gilbert (LLG) equation,[37]

$$\partial m_i/\partial t = -\gamma m_i \times H_{eff} + \alpha m_i \times \partial m_i/\partial t, \quad (1)$$

where i denotes lattice site number, α is the Gilbert damping constant and the effective field arising from different energy terms is given by $H_{eff} = -(1/|\mu_i|) \partial H/\partial m_i$.

MUMAX3, a well-known versatile software is used for simulating the spin dynamics in the structures which incorporate Slonczewski torque due to the flow of current with the Landau-Lifshitz formalism.[14] If only current is present that means only spin transfer torque is active and it will try to align FM spin along input spin current direction (like switching in MTJ). Besides in the definition of spin superfluid, mz (magnetization along Z direction) should be constant (not zero) and to have some mz, magnetic field along Z-axis is necessary. Again, if only magnetic field is present, then there is no force acting to rotate the FM spin. Magnetic field tries to align FM spin along its direction. With the increase in magnetic field along Z-direction, mz will gradually increase. Conditions required for spin superfluid like spin oscillation are the perfect balance between current and magnetic field in materials having easy plane anisotropy.

Magnetization along different directions with oscillations along X-direction and Y-direction for different current density are shown in Fig. 3 using $Fe_3Sn$. Oscillation due to four different magnetic fields is shown in Fig. 4. These figures show spin superfluid oscillation in easy plane (X-Y plane) and frequency increases and again decreases with the increase in both current and magnetic field. Besides increase and decrease in frequency with current and field, several steps in frequency are also found for both CoFeB and $Fe_3Sn$ (Fig. 5). The results are explained below by using the two sublattice model in antiferromagnet or synthetic antiferromagnet.

The directions of the magnetic moments in top and bottom FM of synthetic antiferromagnet are denoted by two-unit vectors m1 and m2. The precession of m1 and m2 are driven by the exchange interaction, the anisotropy, and a magnetic field which is applied along the $\hat{z}$ direction. These three parts are represented by $\omega_E$, $\omega_A$, and $\omega_H = \gamma H_0$, respectively. The equations of motion are [38]

$$\dot{m}_1 = m_1 \times [\omega_E m_2 - (\omega_A + \omega_H)\hat{z}], \quad (2a)$$
$$\dot{m}_2 = m_2 \times [\omega_E m_1 + (\omega_A - \omega_H)\hat{z}], \quad (2b)$$

The resonance frequencies are then
$$\omega = \omega_H \pm \omega_R = \omega_H \pm \sqrt{\omega_A(\omega_A + 2\omega_E)}, \quad (3)$$

and it makes two eigen modes, which are characterized by different chirality. In the absence of magnetic field, ωH = 0, the two modes are degenerate but as magnetic field is present in the simulation, so the two-frequency mode are different. Later damping and torque due to spin current are also incorporated which give rise to the formation of different frequencies with the change in magnetic field and spin current. In antiferromagnet or synthetic antiferromagnet (SAF), there is sizeable difference between two frequency branches of two sublattice due to two frequency mode where these two modes can switch depending on current density [38,39]. This phenomenon gives rise to the variation of frequency with current and magnetic field.

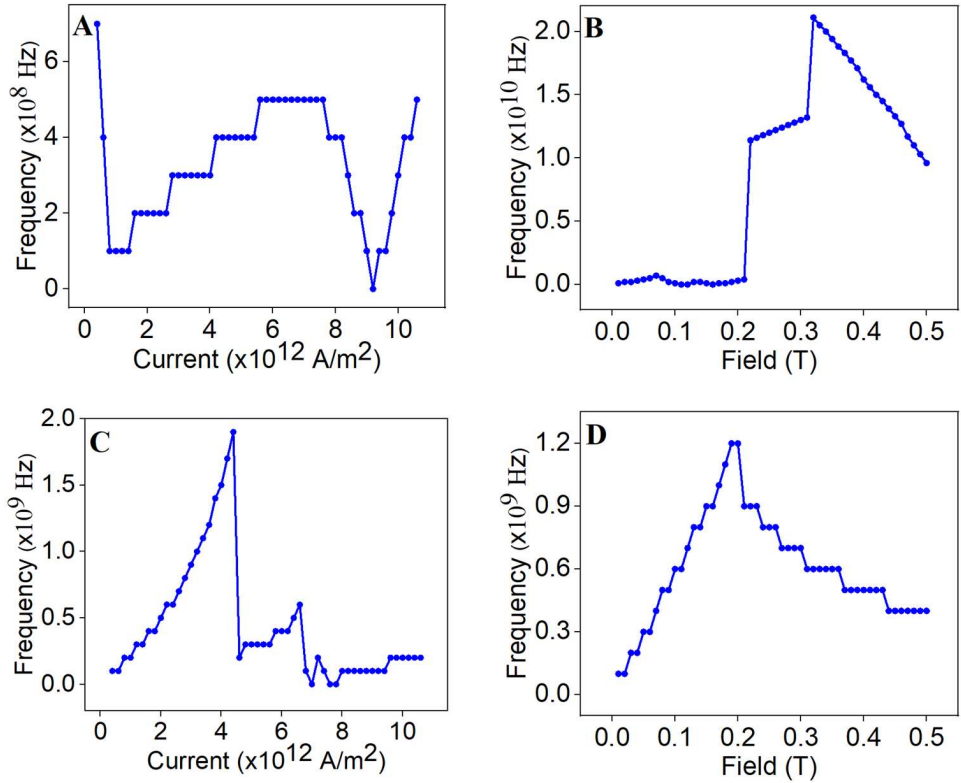

Fig. 5. Variation of spin superfluid oscillation frequency in CoFeB (A) with change in current at applied magnetic field of 0.05T (B) with change in applied magnetic field at current of $5 \times 10^8$ A/cm$^2$. Variation of spin superfluid oscillation frequency in Fe3Sn (C) with change in current at applied magnetic field of 0.05T (D) with change in applied magnetic field at current of $5 \times 10^8$ A/cm2.

## 3. Conclusion

Over the past few decades, research has focused on a wide range of materials and tools that facilitate the production of tiny chips used in a variety of applications [40-53]. Two different materials having easy plane anisotropy are used to host spin superfluidity and SQUID like structures made from these two materials show step-like frequency variation with magnetic field and current. As in neuromorphic computing, synapse requires different weights for different inputs, frequency steps in the proposed device can provide required weights. With the prospective applications in neuromorphic computing and magnetometry, the proposed device can be useful in room temperature applications of SQUID like spin devices.